# Semiconducting transport in $Pb_{10-x}Cu_x(PO_4)_6O$ sintered from $Pb_2SO_5$ and $Cu_3P$

*Li Liu[#], Ziang Meng[#], Xiaoning Wang[#], Hongyu Chen[#], Zhiyuan Duan[#], Xiaorong Zhou, Han Yan, Peixin Qin\*, Zhiqi Liu\**

School of Materials Science and Engineering, Beihang University, Beijing 100191, China

#These authors contributed equally to this work.

Emails:qinpeixin@buaa.edu.cn; zhiqi@buaa.edu.cn

**The very recent claim on the discovery of ambient-pressure room-temperature superconductivity in modified lead-apatite has immediately excited sensational attention in the entire society, which is fabricated by sintering lanarkite ($Pb_2SO_5$) and copper(I) phosphide ($Cu_3P$). To verify this exciting claim, we have successfully synthesized $Pb_2SO_5$, $Cu_3P$, and finally the modified lead-apatite $Pb_{10-x}Cu_x(PO_4)_6O$. Detailed electrical transport and magnetic properties of these compounds were systematically analyzed. It turns out that $Pb_2SO_5$ is a highly insulating diamagnet with a room-temperature resistivity of $\sim 7.18\times 10^9$ $\Omega\cdot$cm and $Cu_3P$ is a paramagnetic metal with a room-temperature resistivity of $\sim 5.22\times 10^{-4}$ $\Omega\cdot$cm. In contrast to the claimed superconductivity, the resulting $Pb_{10-x}Cu_x(PO_4)_6O$ compound sintered from $Pb_2SO_5$ and $Cu_3P$ exhibits semiconductor-like transport behavior with a large room-temperature resistivity of $\sim 1.94\times 10^4$ $\Omega\cdot$cm although our compound shows greatly consistent x-ray diffraction spectrum with the previously reported structure data. In addition, when a pressed $Pb_{10-x}Cu_x(PO_4)_6O$ pellet is located on top of a commercial $Nd_2Fe_{14}B$ magnet at room temperature, no repulsion could be felt and no magnetic levitation was observed either. These results imply that the claim of a room-temperature superconductor in modified lead-apatite may need more careful re-examination, especially for the electrical transport properties.**

## 1. Introduction

Superconductivity depicts the zero-resistance state of a material in physics, which has been a center of focus for materials science and condensed matter physics for more than one century. Meanwhile, searching for high-temperature superconductors has been always one of the major goals in these fields. It is rather straightforward that once room-temperature superconductivity could be realized, the electrical energy consumption for the entire human being society would be significantly reduced as the Joule heating would be largely suppressed. Accordingly, without doubt, the realization of room-temperature superconductivity would completely revolutionize the way how the people live on the earth.

There have reports on room-temperature superconductors, such as nitrogen-doped lutetium hydride with a maximum superconducting transition temperature $T_c$ of 294 K at 10 kbar.[1] However, to achieve the superconductivity, a hydrostatic pressure is needed, which prevents these materials from practical applications. More importantly, it is still rather controversial.[2] Very recently, the claim of ambient-pressure room-temperature superconductivity for a modified lead-apatite[3,4] has promptly motivated a dramatic surge of research interest in synthesizing such a material. These reports[3,4] are exceptionally exciting as the superconductivity was reported to occur at ambient pressure and at temperatures even higher than 300 K, the synthesis process of the material system seems to be quite straightforward, and the demonstrated magnetic levitation phenomenon is rather attractive.

In this work, following the reported fabrication procedure,[3,4] we have tried to synthesize the lanarkite $Pb_2SO_5$, copper(I) phosphide $Cu_3P$, and the modified lead-apatite $Pb_{10-x}Cu_x(PO_4)_6O$ with great interest and curiosity. Subsequently, electrical transport and magnetic properties of these materials were carefully determined by four-probe resistivity and magnetometry measurements.

## 2. Synthesis, electrical transport and magnetic properties of lanarkite $Pb_2SO_5$

As reported by Lee *et al.*,[3] the lanarkite $Pb_2SO_5$ can be straightforwardly fabricated by sintering lead(Ⅱ) sulfate $PbSO_4$ and lead oxide $PbO$ with a molar ration of 1:1 at 725°C for 24 h. Such a process was conductor in our own lab. The step-by-step synthesis process is shown in **Figure 1**. After the sintering, the resulting powder was confirmed to be purely the lanarkite $Pb_2SO_5$ phase by x-ray diffraction analysis.

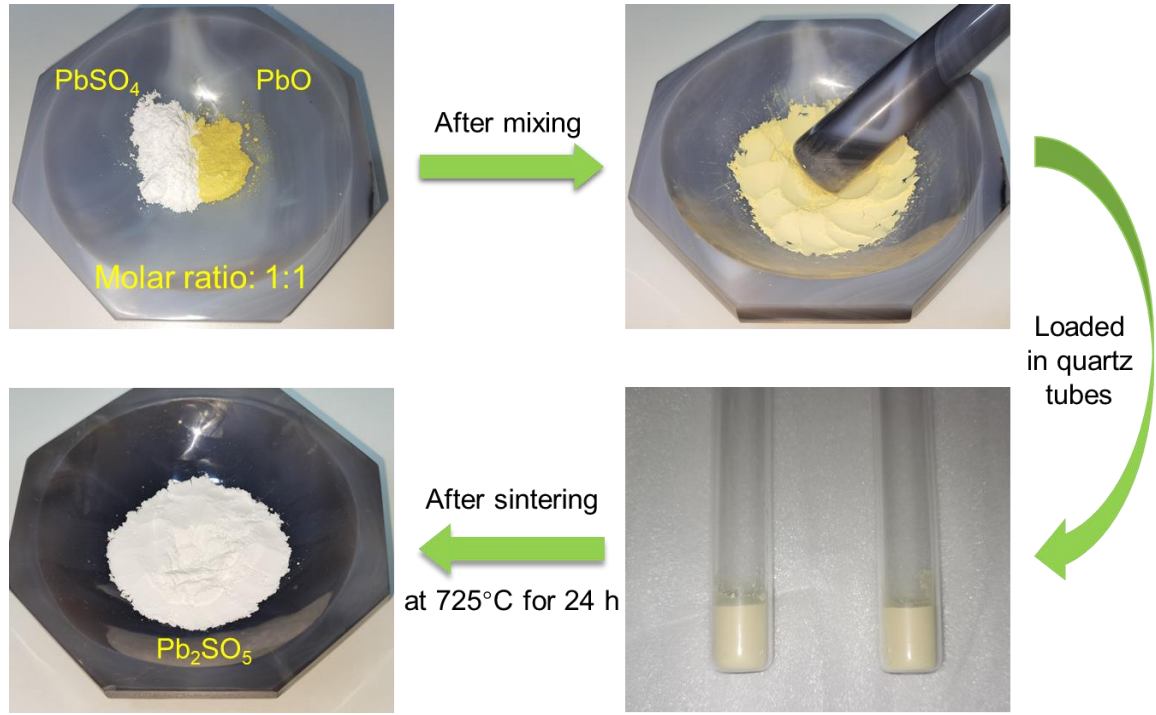

**Figure 1.** Step-by-step synthesis process of $Pb_2SO_5$.

To investigate the electrical properties of this material, the sintered and ground $Pb_2SO_5$ powder was tableted into a pellet. Furthermore, the pellet was mounted onto a standard Quantum Design physical properties measurement system puck and linear four-probe electrical contacts (**Figure 2**a) were made by Al wires via a wire bond system and silver paint. It was found that such a pellet is highly insulating. The room-temperature current-voltage (*I-V*) curve plotted in Figure 2b reveals a huge resistivity of ~$7.18\times10^9$ Ω·cm, characteristic of an insulator. This is consistent with the large theoretically calculated band gap, ~3.0 eV.[5] In addition, magnetometry measurements on $Pb_2SO_5$ powder (**Figure 3**a) conducted by a Quantum Design magnetometer with a magnetic moment sensitivity of ~$10^{-7}$ emu indicate that it is an ideal diamagnet with temperature-independent negative magnetization (Figure 3b). The magnetization is ~$-10^{-4}$ emu/g at a magnetic field of 0.5 T, which is typical for an oxide.

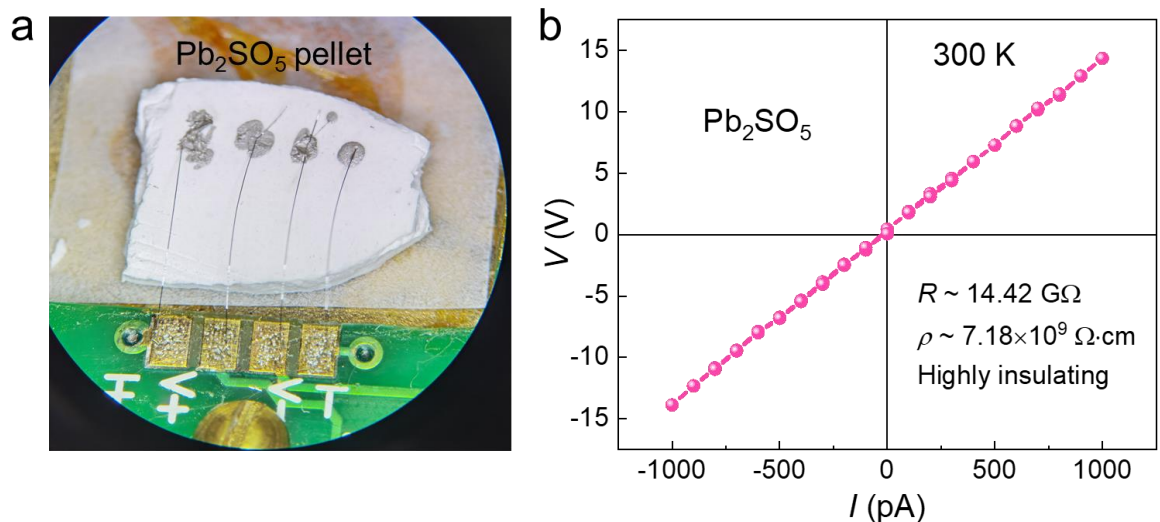

**Figure 2.** Room-temperature resistivity of a $Pb_2SO_5$ pellet. a) Linear four-probe electrical contacts made by wire bond and silver paint. b) Room-temperature four-probe $I$-$V$ characteristic measured up to 1 nA.

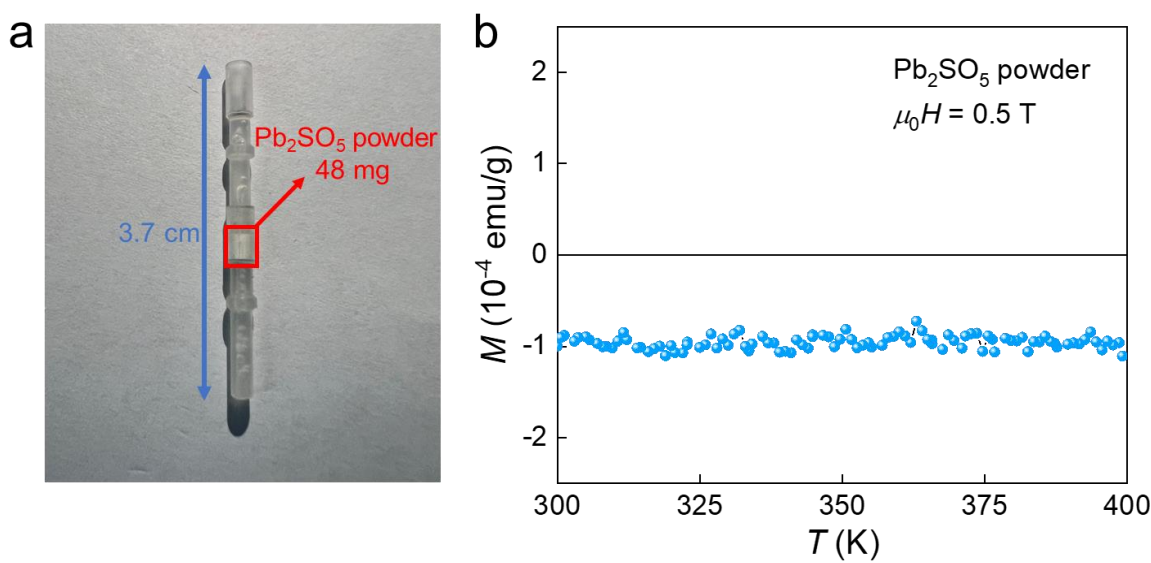

**Figure 3.** Magnetization of $Pb_2SO_5$ powder. a) Loaded $Pb_2SO_5$ powder in a Quantum Design magnetometer capsule. The mass of the powder was measured to be ~48 mg. b) Magnetization versus temperature measured by a 0.5 T magnetic field between 300 and 400 K.

## 3. Synthesis, electrical transport and magnetic properties of copper(I) phosphide $Cu_3P$

The synthesis procedure of $Cu_3P$ was reported[3] to be sintering the Cu and red phosphorus powder (molar ratio of 3:1) in a vacuum tube at 550°C for 48 h. Accordingly, we have performed this procedure conscientiously and the step-by-step process is presented in **Figure 4**. Similarly, the sintered product was verified by powder x-ray diffraction to be purely the $Cu_3P$ phase, which is consistent with the results reported by Lee *et al.*[3,4]

To explore the electrical transport properties, $Cu_3P$ powder was pressed into a pellet and the linear four-probe electrical contacts were established solely by an Al wire bond system (**Figure 5**a). Room-temperature $I$-$V$ characteristic (Figure 5b) measurements indicate that the synthesized $Cu_3P$ pellet is greatly conductive material with a resistivity of $\sim 5.22\times 10^{-4}$ $\Omega\cdot cm$, which is comparable with the room-temperature resistivity of some intermetallic compounds such as FeRh, $Mn_3Sn$ and Mn-Pt systems.[6-9]

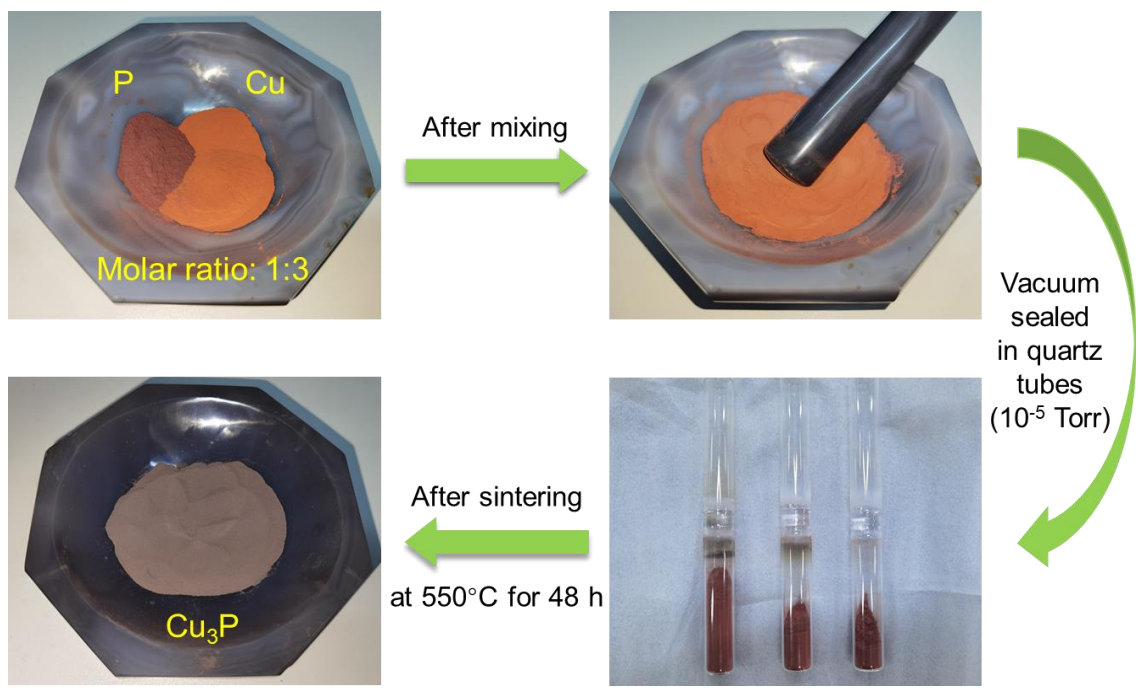

**Figure 4.** Step-by-step synthesis process of $Cu_3P$.

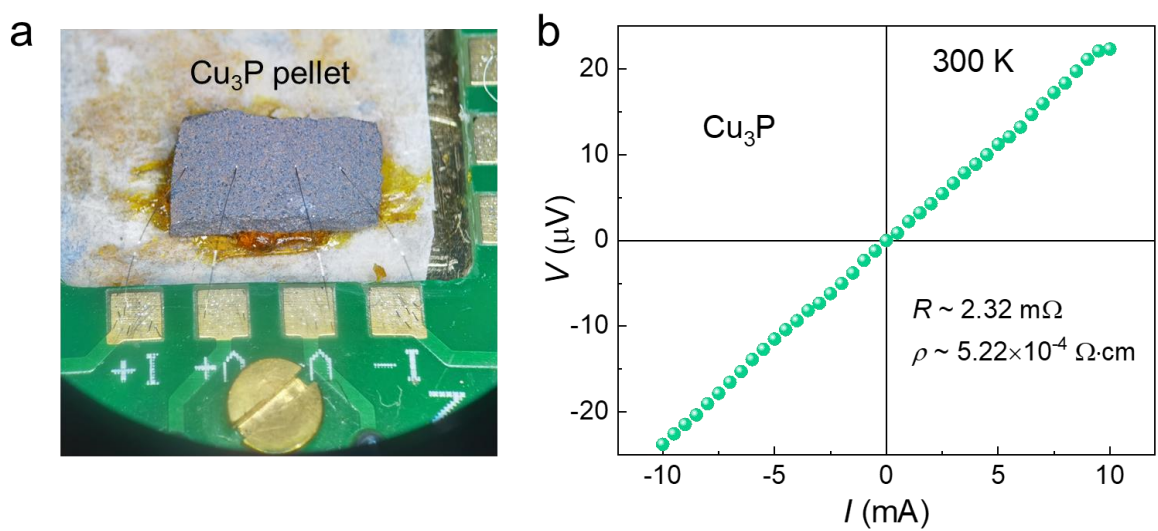

**Figure 5.** Room-temperature resistivity of a $Cu_3P$ pellet. a) Linear four-probe electrical contacts constructed by an Al wire bond system. b) Room-temperature four-probe *I-V* characteristic measured up to 10 mA.

As the pressed $Cu_3P$ pellet is a conducting system, detailed electrical transport properties were further examined. As shown in **Figure 6**a, the temperature-dependent resistivity between 400 and 50 K exhibits classical metallic behavior and a linear decrease of the resistivity with lowering temperature is clearly seen, which demonstrates that our synthesized $Cu_3P$ pellet is a metal. Due to the large carrier density, the Hall resistance measurements up to 3 T were not able to yield reliable data points upon varying the external magnetic field. Considering the measurement limit of our combined Keithley 6221/2182A meters, the estimated carrier density is larger than $10^{22}$ cm$^{-3}$, which is typical for a metal. This is largely distinct from the semimetal behavior of the recently fabricated Cu$_3$P thin films.[10] Room-temperature resistance versus magnetic field plotted in Figure 6b indicates that the magnetoresistance effect in Cu$_3$P is negligible. In addition, the magnetization measurements (**Figure 7**) performed between 200 and 400 K unveil that Cu$_3$P is a prototypical paramagnet with a Curie-Weiss type magnetism.

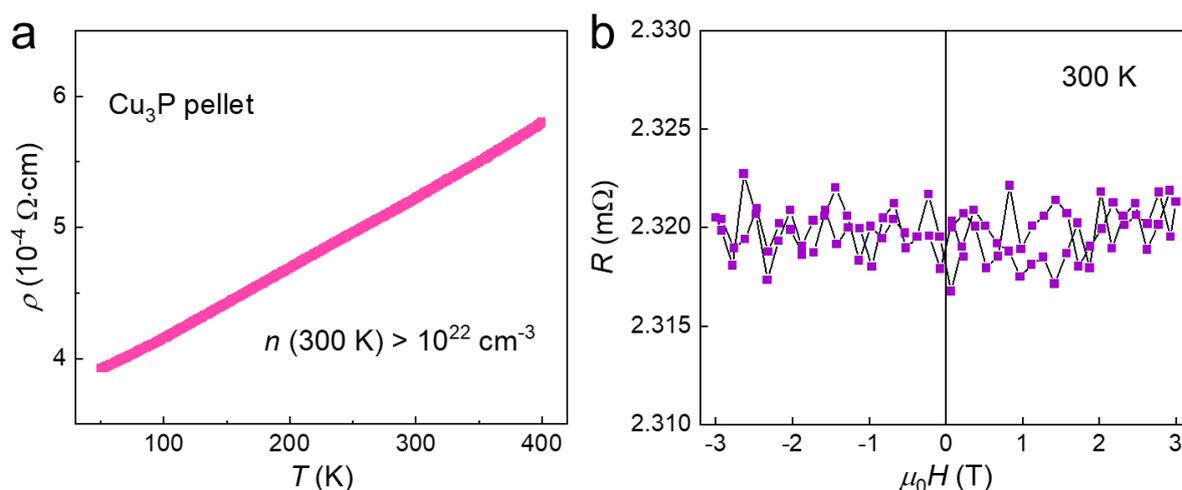

**Figure 6.** Electrical transport behavior of the Cu$_3$P pellet. a) Temperature-dependent resistivity from 400 to 50 K. b) Resistance as a function of magnetic field at room temperature.

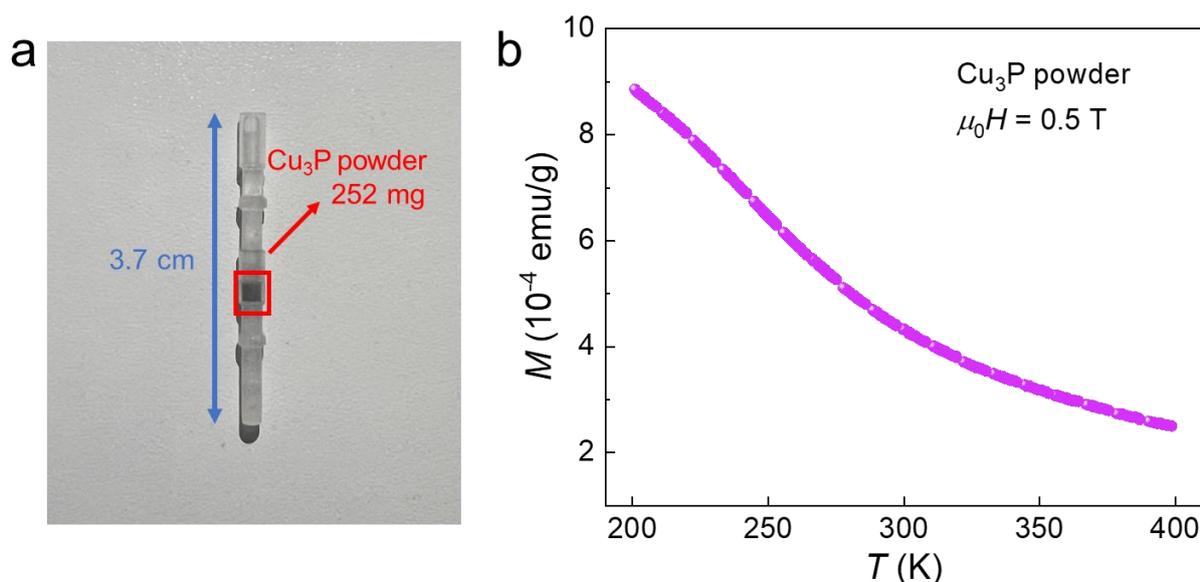

**Figure 7.** Magnetization of Cu$_3$P powder. a) Loaded Cu$_3$P powder in a Quantum Design magnetometer capsule. The mass of the powder was measured to be ~252 mg. b) Magnetization versus temperature measured by a 0.5 T magnetic field between 200 and 400 K.

# 4. Synthesis, structural, electrical transport and magnetic properties of modified lead-apatite

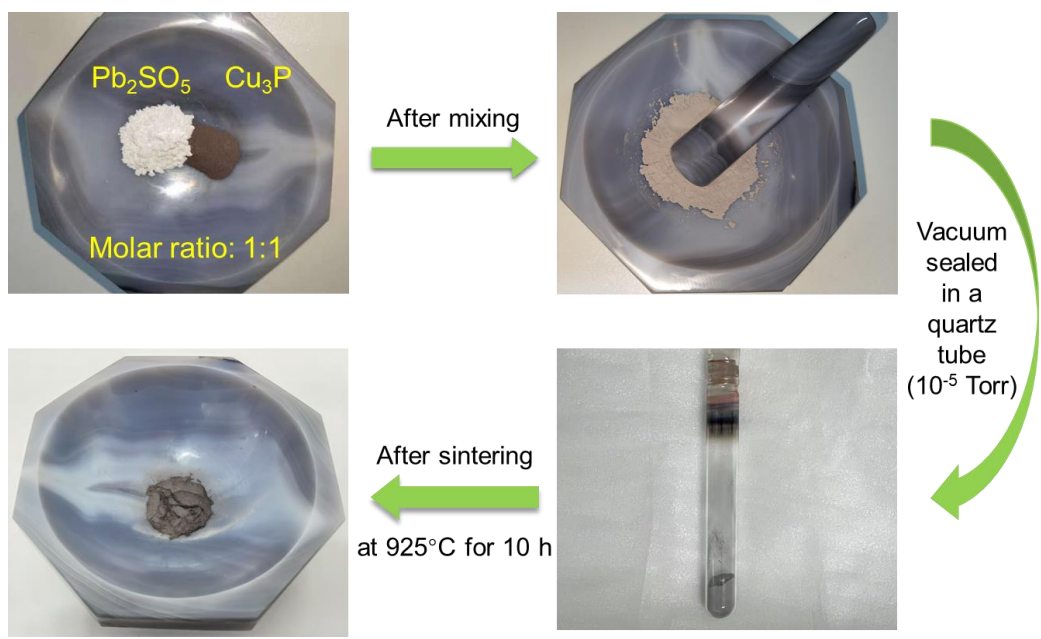

**Figure 8.** Step-by-step procedure for the synthesis of the modified lead-apatite $Pb_{10-x}Cu_x(PO_4)_6O$.

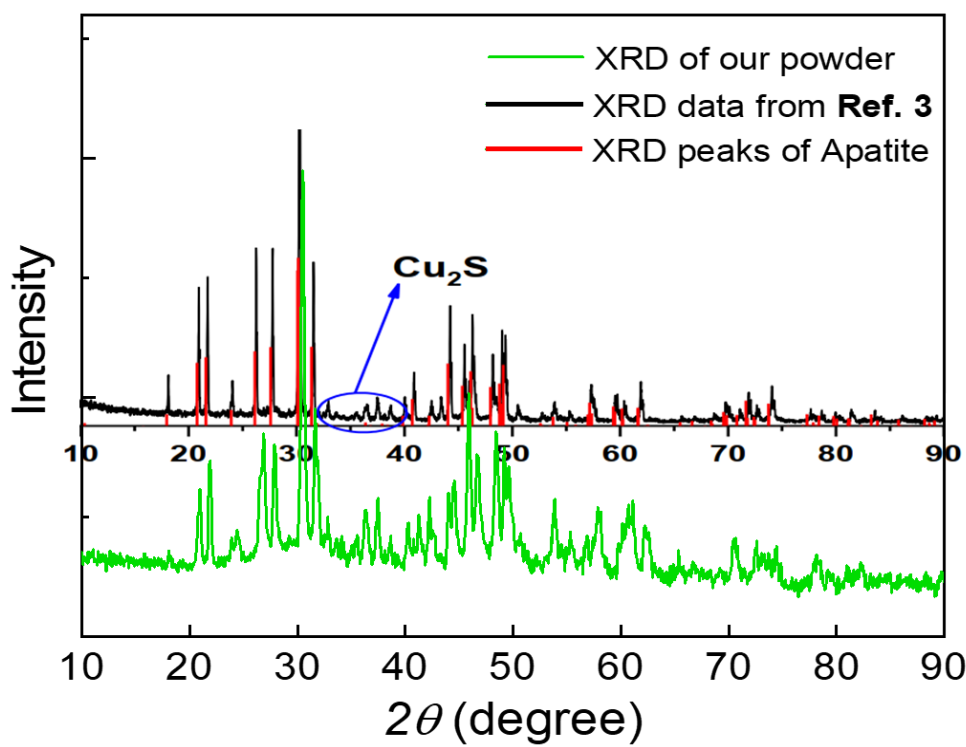

**Figure 9.** X-ray diffraction spectrum (green) of our sintered powder and other reference data. The black spectrum and the red lines are adapted from the report by Lee *et al*.[3]

The reported modified lead-apatite was straightforwardly synthesized by sintering the lanarkite $Pb_2SO_5$ and the copper(I) phosphide $Cu_3P$ (molar ratio: 1:1) in a sealed vacuum tube at 925°C for 10 h. We have followed this routine to fabricate the final desired product. **Figure 8** depicts our laboratory synthesis procedure. As shown in **Figure 9**, the x-ray diffraction spectrum of the ground powder of the finally sintered product is highly consistent with the x-ray diffraction spectrum reported by Lee *et al*.[3] and coincides well with the diffraction pattern of the apatite. This proves that we have successfully synthesized the modified lead-apatite as Lee *et al*.[3,4]

To demonstrate the superconductivity for a new material, the most convincing evidence shall be the steady and low-noise zero-resistance state, which in practical electrical measurements may be reflected by the lowest resistance reaching the measuring sensitivity of the utilized electronic equipment combined with source/sensing meters. To investigate whether the newly synthesized $Pb_{10-x}Cu_x(PO_4)_6O$ contains a superconducting state, its powder was tabletted into dense pellets for electrical resistance measurements. As shown in **Figure 10**, the linear four-probe *I-V* characteristic collected at room temperature reveals a resistivity of ~$1.94\times10^4$ $\Omega\cdot$cm. This is at least 7 orders of magnitude greater than the room-temperature resistivity of metals as most of metallic materials possess a room-temperature resistivity smaller than $10^{-3}$ $\Omega\cdot$cm.[11] Therefore, it implies that $Pb_{10-x}Cu_x(PO_4)_6O$ is a semiconductor rather than a metal.

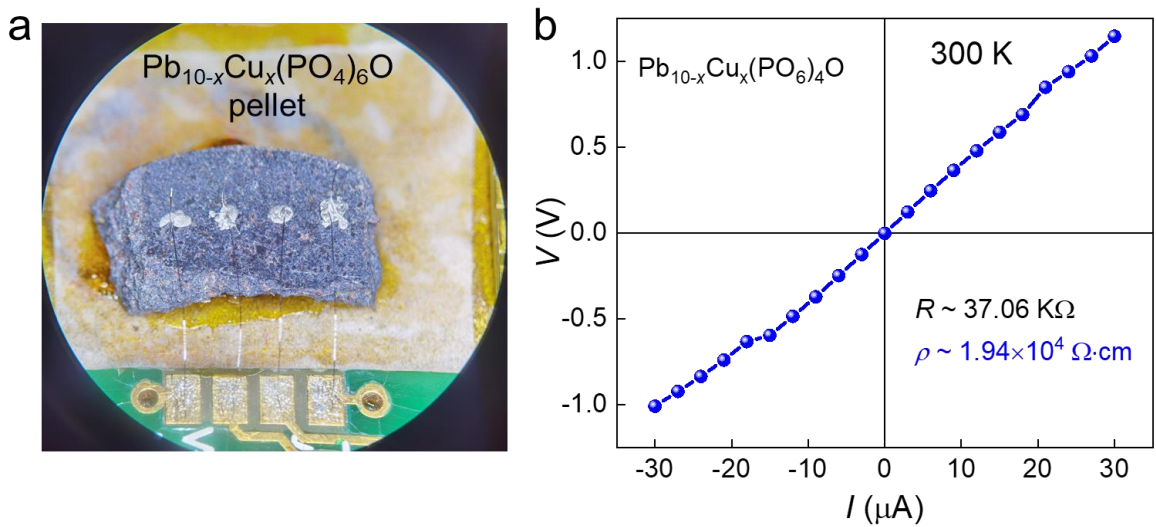

**Figure 10.** Room-temperature resistivity of a $Pb_{10-x}Cu_x(PO_4)_6O$ pellet. a) Linear four-probe electrical contacts made by wire bond and silver paint. b) Room-temperature four-probe *I-V* characteristic measured up to 30 μA.

To further explore the electrical transport behavior, temperature-dependent resistivity was measured for the $Pb_{10-x}Cu_x(PO_4)_6O$ pellet. As shown in **Figure 11**, the resistivity of dramatically increases with decreasing temperature. The temperature-dependent resistivity is almost linear on a semi-logarithmic scale, characteristic of a prototypical semiconductor. This corroborates the semiconducting transport behavior of the modified lead-apatite. This is in sharp contrast to the claimed "zero resistivity" by Lee *et al*.[3,4] We have noticed that there seem contradictory claims in the electrical resistivity data in the two reports. For example, in the first report,[3] the fluctuation sensing voltage noise level for the "zero resistivity" is ~0.1 μV, while it is ~0.1 mV[4] in the second report. In addition, the low-resistance state (referred as the "superconducting state") in the second report by Lee *et al.*[4] is ~$10^{-3}$ Ω·cm, which is comparable with the room-temperature resistivity of a usual metallic material and may have rare connection with superconductivity.

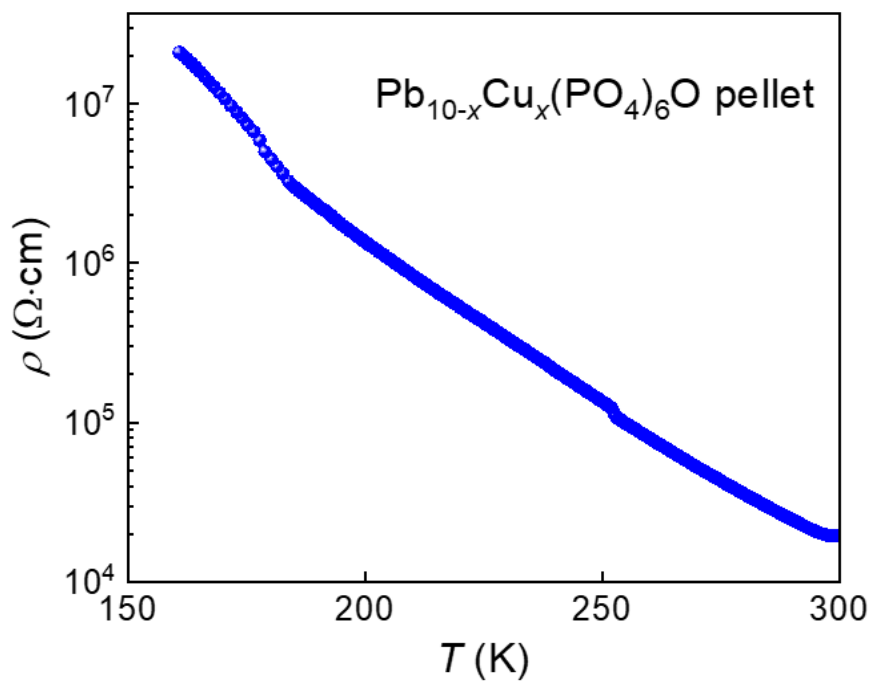

**Figure 11.** Temperature-dependent resistivity for the tableted $Pb_{10-x}Cu_x(PO_4)_6O$ pellet.

It is still confusing that our electrical transport data are largely different from the previous "zero resistivity" reports[3,4] although our material exhibits a highly similar structure with Lee *et al.*'s materials. Nevertheless, it is worth emphasizing there could be many measurement artifacts when dealing with electrical transport properties of oxides. That is because the electrical contacts between metal electrodes and oxides typically contain Schottky junctions, and any poor contacts can result in large contract resistance, which, in turn, might yield "zero resistance " artifact while the intrinsic resistivity could be much higher and could even exceed the upper resistance measurement limit of the equipment.

Another highly interesting aspect of the report of Lee *et al.*[3] is the large diamagnetism, ~-7.4 ×$10^{-4}$ emu/g at a small measuring magnetic field of 1 mT (10 Oe). To explore this aspect, we have also performed the zero-field cooling magnetization measurements for the $Pb_{10-x}Cu_x(PO_4)_6O$ powder. Unfortunately, under a 1 mT measuring field, we could not detect any reliable diamagnetic signal within our measurement sensitivity of $10^{-7}$ emu, which suggests that under 1 mT, the diamagnetic magnetization of our powder is smaller than -1.61×$10^{-6}$ emu/g.

This is two orders of magnitude smaller than the giant diamagnetism reported for $Pb_{10-x}Cu_x(PO_4)_6O$ by Lee *et al.*[3] At a large magnetic field of 0.5 T, our $Pb_{10-x}Cu_x(PO_4)_6O$ powder exhibit a paramagnetic behavior (**Figure 12**). Moreover, when a tableted $Pb_{10-x}Cu_x(PO_4)_6O$ pellet is placed on top of a commercial $Nd_2Fe_{14}B$ magnet, we could not feel any repulsion and no magnetic levitation was observed either (**Figure 13**).

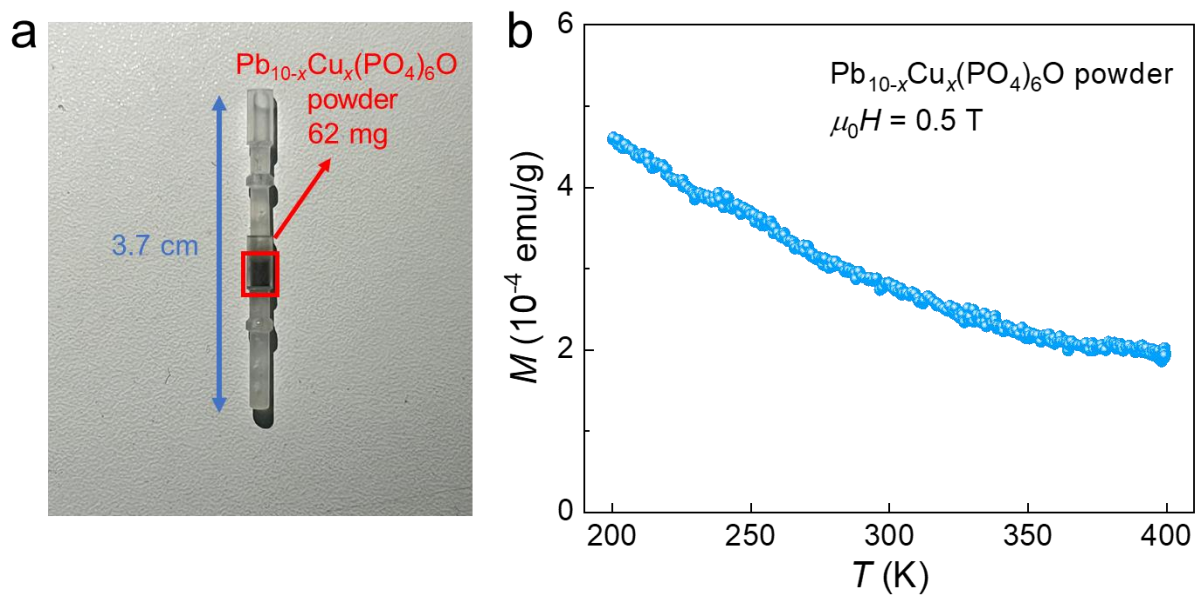

**Figure 12.** Magnetization of $Pb_{10-x}Cu_x(PO_4)_6O$ powder. a) Loaded $Pb_{10-x}Cu_x(PO_4)_6O$ powder in a Quantum Design magnetometer capsule. The mass of the powder was measured to be ~62 mg. b) Magnetization versus temperature measured by a 0.5 T magnetic field between 200 and 400 K.

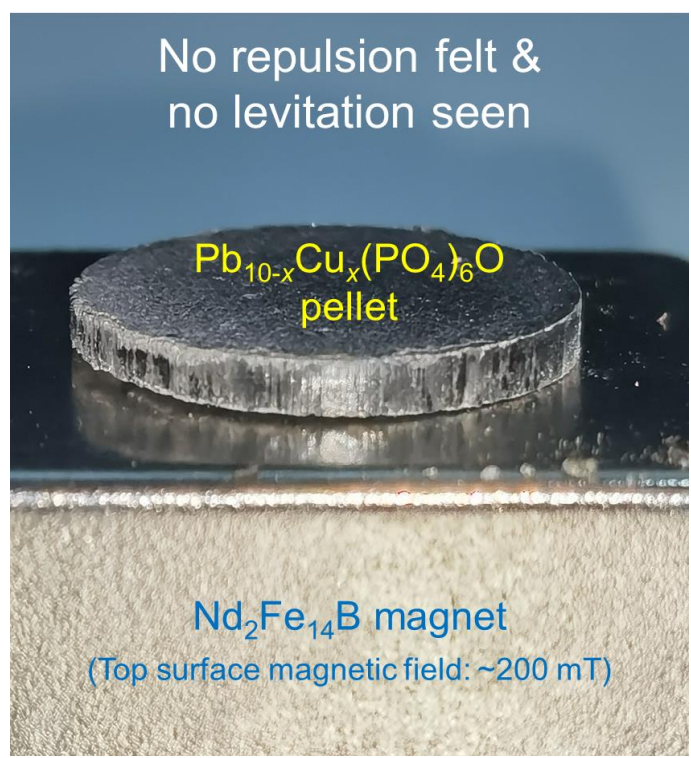

**Figure 13.** No repulsion felt and no magnetic levitation observed when a tableted pellet is placed on top of a commercial $Nd_2Fe_{14}B$ permanent magnet with a top surface magnetic field of ~200 mT.

## 5. Summary

To briefly summarize, as motivated by the highly exciting claim on the ambient-pressure room-temperature superconductivity in modified lead-apatite, we have successfully synthesized the lanarkite $Pb_2SO_5$, the copper(I) phosphide $Cu_3P$, and the modified lead-apatite $Pb_{10-x}Cu_x(PO_4)_6O$. Electrical and magnetic properties of these materials have been comprehensively examined. Our results show that the modified lead-apatite $Pb_{10-x}Cu_x(PO_4)_6O$ is semiconducting with a large room temperature resistivity on the order of $10^4\,\Omega\cdot cm$, rather than superconducting. These results suggest that the claim of room-temperature resistivity in $Pb_{10-x}Cu_x(PO_4)_6O$ needs more careful re-examination and investigation, especially from the reliable electrical transport perspective. We firmly believe that more and more experimental results will soon appear from other groups over the world and hope our own experimental results in this work could be helpful to clarify the exciting reports.

**Acknowledgements**
Z.L. acknowledges the financial support of the National Key R&D Program of China (Grant No. 2022YFB3506000). Z.L. acknowledges the financial support of the National Key R&D Program of China (Grant No. 2022YFA1602701). Z.L. acknowledges the funding supported by the National Natural Science Foundation of China (Nos. 52271235 and 52121001).